\documentclass[fleqn,11pt,twoside]{article}

\usepackage{amsthm,amssymb, color, epsfig, graphics, subfigure}

\usepackage{tikz}

\usepackage{graphicx}
\usepackage{latexsym}
\usepackage[all]{xy}
\usepackage{amsmath, amscd}
\usepackage{graphicx}
\usepackage{lscape}

\newtheorem{proposition}{Proposition}

\newtheorem{conjecture}{Conjecture}

\makeatletter
\newcommand{\copyrightnote}[2]{{\renewcommand{\thefootnote}{}
 \footnotetext{\small\it
\begin{flushleft}
\copyright \ #1 
\end{flushleft}}}}

\newcommand{\Name}[1]{\begin{flushleft}
                       \LARGE \bf #1
                       \end{flushleft}\vspace{-3mm}}

\newcommand{\Author}[1]{\begin{flushleft}
                       \it #1 \end{flushleft}}

\newcommand{\Address}[1]{\begin{flushleft}
                       \it #1 \end{flushleft}}

\newcommand{\Date}[1]{\begin{flushleft}
                      \small  \it #1 \end{flushleft}}

\newcommand{\evenhead}{Author \ name}
\newcommand{\oddhead}{Article \ name}

\renewcommand{\@evenhead}{
\hspace*{-3pt}\raisebox{-15pt}[\headheight][0pt]{\vbox{\hbox to \textwidth
{\thepage \hfil \evenhead}\vskip4pt \hrule}}}
\renewcommand{\@oddhead}{
\hspace*{-3pt}\raisebox{-15pt}[\headheight][0pt]{\vbox{\hbox to \textwidth
{\oddhead \hfil \thepage}\vskip4pt\hrule}}}
\renewcommand{\@evenfoot}{}
\renewcommand{\@oddfoot}{}

\long\def\@makecaption#1#2{%
  \vskip\abovecaptionskip
  \sbox\@tempboxa{\small \textbf{#1.}\ \ #2}%
  \ifdim \wd\@tempboxa >\hsize
    {\small \textbf{#1.}\ \ #2}\par
  \else
    \global \@minipagefalse
    \hb@xt@\hsize{\hfil\box\@tempboxa\hfil}%
  \fi
  \vskip\belowcaptionskip}

\newcommand{\JNMPnumberwithin}[3][\arabic]{%
  \@ifundefined{c@#2}{\@nocounterr{#2}}{%
    \@ifundefined{c@#3}{\@nocnterr{#3}}{%
      \@addtoreset{#2}{#3}%
      \@xp\xdef\csname the#2\endcsname{%
        \@xp\@nx\csname the#3\endcsname .\@nx#1{#2}}}}%
}

\newcommand{\resetfootnoterule} {
  \renewcommand\footnoterule{%
  \kern-3\p@
  \hrule\@width.4\columnwidth
  \kern2.6\p@}
}

\renewcommand{\footnoterule}{}

\newcommand{\be}{\begin{equation}}
\newcommand{\ee}{\end{equation}}
\newcommand{\ba}{\hspace*{-5pt}\begin{array}}
\newcommand{\ea}{\end{array}}
\newcommand{\p}{\partial}

\makeatother

\numberwithin{equation}{section}
\theoremstyle{definition}

\theoremstyle{proposition}
\theoremstyle{conjecture}

\renewcommand{\ba}{\begin{array}}
\renewcommand{\ea}{\end{array}}
\newcommand{\beg}{\begin{eqnarray}}
\newcommand{\eeq}{\end{eqnarray}}
\newcommand{\bg}{\begin{eqnarray*}}

\newcommand{\ed}{\end{eqnarray*}}

\newcommand{\nn}{\nonumber}

\renewcommand{\p}{\partial} 
 
\newcommand{\notlhd}{\lhd\kern-.8em{/}\ } 
\newcommand{\notexist}{\ \exists\kern-.5em{\raise.1em\hbox{/}}\ }

\newcommand{\pde}[2]{\frac{\p #1}{\p #2}}

\newcommand{\inp}{{\mbox{\vbox{\hrule width0ex\hbox{\vrule
 height0ex\kern3.8pt
\vbox{\kern2.5pt}\kern3.8pt \vrule height1.6ex}
\hrule width1.6ex}}}}

\makeatletter
\newcommand*{\da@rightarrow}{\mathchar"0\hexnumber@\symAMSa 4B }
\newcommand*{\da@leftarrow}{\mathchar"0\hexnumber@\symAMSa 4C }
\newcommand*{\xdashrightarrow}[2][]{%
  \mathrel{%
    \mathpalette{\da@xarrow{#1}{#2}{}\da@rightarrow{\,}{}}{}%
  }%
}
\newcommand{\xdashleftarrow}[2][]{%
  \mathrel{%
    \mathpalette{\da@xarrow{#1}{#2}\da@leftarrow{}{}{\,}}{}%
  }%
}
\newcommand*{\da@xarrow}[7]{%
  \sbox0{$\ifx#7\scriptstyle\scriptscriptstyle\else\scriptstyle\fi#5#1#6\m@th$}%
  \sbox2{$\ifx#7\scriptstyle\scriptscriptstyle\else\scriptstyle\fi#5#2#6\m@th$}%
  \sbox4{$#7\dabar@\m@th$}%
  \dimen@=\wd0 %
  \ifdim\wd2 >\dimen@
    \dimen@=\wd2 %
  \fi
  \count@=2 %
  \def\da@bars{\dabar@\dabar@}%
  \@whiledim\count@\wd4<\dimen@\do{%
    \advance\count@\@ne
    \expandafter\def\expandafter\da@bars\expandafter{%
      \da@bars
      \dabar@ 
    }%
  }%
  \mathrel{#3}%
  \mathrel{%
    \mathop{\da@bars}\limits
    \ifx\\#1\\%
    \else
      _{\copy0}%
    \fi
    \ifx\\#2\\%
    \else
      ^{\copy2}%
    \fi
  }%
  \mathrel{#4}%
}
\makeatother

\begin{document}

\renewcommand{\evenhead}{ {\LARGE\textcolor{blue!10!black!40!green}{{\sf \ \ \ ]ocnmp[}}}\strut\hfill M Euler and N Euler}
\renewcommand{\oddhead}{ {\LARGE\textcolor{blue!10!black!40!green}{{\sf ]ocnmp[}}}\ \ \ \ \   
Two sequences of fully-nonlinear evolution equations: symmetries
}

\thispagestyle{empty}
\newcommand{\FistPageHead}[3]{
\begin{flushleft}
\raisebox{8mm}[0pt][0pt]
{\footnotesize \sf
\parbox{150mm}{{Open Communications in Nonlinear Mathematical Physics}\ \ \ {\LARGE\textcolor{blue!10!black!40!green}{]ocnmp[}}
\ \ Vol.5 (2025) pp
#2\hfill {\sc #3}}}\vspace{-13mm}
\end{flushleft}}

\FistPageHead{1}{\pageref{firstpage}--\pageref{lastpage}}{ \ \ Letter}

\strut\hfill

\strut\hfill

\copyrightnote{The author(s). Distributed under a Creative Commons Attribution 4.0 International License}

\qquad\qquad\qquad\qquad\qquad\qquad {\LARGE  {\sf Letter to the Editors}}

\strut\hfill

\Name{Two sequences of fully-nonlinear evolution equations and their symmetry properties}

\Author{Marianna Euler and Norbert Euler$^{\,*}$}

\Address{International Society of Nonlinear Mathematical Physics (ISNMP)\\ 
Auf der Hardt 27, 56130 Bad Ems, Germany\\
and\\
Centro Internacional de Ciencias (CIC AC)\\ 
Av. Universidad s/n, Colonia Chamilpa,
62210 Cuernavaca, Morelos, Mexico\\[-0.2cm]
\ \\
 $^*$ Dr.Norbert.Euler@gmail.com }

\Date{Received September 9, 2025; Accepted October 10, 2025}

\setcounter{equation}{0}

\begin{abstract}
\noindent 
We obtain the complete Lie point symmetry algebras of two sequences of odd-order evolution equations. This includes equations that are fully-nonlinear, i.e. nonlinear in the highest derivative. Two of the equations in the sequences have recently been identified as symmetry-integrable, namely a 3rd-order equation and a 5th-order equation [{\it Open Communications in Nonlinear Mathematical Physics}, {\it Special Issue in honour of George W Bluman}, ocnmp:15938, 1--15, 2025]. These two examples provided the motivation for the current study. The Lie-Bäcklund symmetries and the consequent symmetry-integrability of the equations in the sequences are also discussed.
\end{abstract}

\label{firstpage}


\section{Introduction}

We recently \cite{Euler-Euler-2025-Bluman} reported two fully-nonlinear symmetry-integrable evolution equations, namely the 3rd-order equation
\begin{gather}
\label{3rd-order-example}
u_t=u_{3x}^{-1/2}
\end{gather}
and the 5th-order equation
\begin{gather}
\label{5th-order-example}
u_t=u_{5x}^{-2/3}.
\end{gather}
Througout this Letter we make use of the notation $u_x=\p u/\p x$, $u_{xx}:=\p^2 u/\p x^2$ and $u_{px}:=\p^p u/\p x^p$ for $p\geq 3$.  
Motivated by the symmetry-integrability of equations (\ref{3rd-order-example}) and (\ref{5th-order-example}), we propose here the following sequence of odd-order evolution equations:
\begin{gather}
\label{gen-hier}
u_t=\left(u_{(2k+1)x}\right)^{-n_k},\quad k=1,2,3,...,
\end{gather}
that is
\begin{gather}
u_t=u_{3x}^{-n_1},\quad
u_t=u_{5x}^{-n_2},\quad 
u_t=u_{7x}^{-n_3},
\ldots\ .
\end{gather}
Here $n_k$ is a number which is, in general, different for every value of $k$ and $n_k\notin\{ -1,0\}$ for any $k$.
It will be shown in Sections 2 and Section 3 that the relation between $n_k$ and $k$ is essential for the symmetry properties and the symmetry-integrability of the equations in this sequence. We note that equations (\ref{3rd-order-example}) and 
(\ref{5th-order-example}) are included as the first two members of the sequence (\ref{gen-hier}) for the case where
$n_k={k}/(k+1)$, namely $k=1$ and $k=2$, respectively. 

Note further that we do not consider here sequences of even order of the form 
\begin{gather}
\label{even}
u_t=\left(u_{2kx}\right)^{n_k},\quad k=1,2,3,\ldots
\end{gather}
since the 2nd-order case (so (\ref{even}) for $k=1$) is included in our study of general 2nd-order evolution equations \cite{E-E-Conf-2024})
in which we have established that the equation
\begin{gather}
u_{t}=u_{xx}^{-1},
\end{gather}
is the only symmetry-integrable equation of the form (\ref{even}).  In particular, (\ref{even}) admits the Lie-Bäcklund symmetry generator
\begin{gather}
Z_{LB}=u_{xx}^{-3} u_{3x}\, \pde{\ }{u}.
\end{gather}
All other even-order equations of the form (\ref{even}), i.e. for $k\geq 2$, do not satisfy the necessary conditions for symmetry-integrability (see Conjecture 1 in Section 3 below).

In this short Letter we derive the Lie point symmetry algebras of (\ref{gen-hier}) in Section 2. The symmetry analysis reveals that 
(\ref{gen-hier}) in fact consists of two sequences, whereby the difference between the two sequences is essentially given by one symmetry. We furthermore discuss the Lie-Bäcklund symmetry structure of the two sequences in Section 3, i.e. the symmetry-integrability of the equations, for which we state a Conjecture with a Corollary. In Section 4 we then sum up our findings and make some concluding remarks.

\section{The Lie point symmetries of the sequences}

In this section we report the Lie point symmetries of the sequence (\ref{gen-hier}). It is well-known how to calculate Lie point symmetries of evolution equations (see for example the books \cite{Olver} or \cite{Bluman+Anco}) so we merely point out our notations here (see \cite{Euler-book-2018} for details).

\strut\hfill

\noindent
Let $E(x,t,u,u_t,u_x,u_{xx},\ldots,u_{nx})=0$ denote a general $n$th-order partial differential equation. For
evolution equations we have
\begin{gather}
\label{Evo-Eq}
E:=u_t-F(x,t,u,u_x,u_{xx},\ldots,u_{nx}).
\end{gather}
The invariance condition for the Lie point symmetries of  (\ref{Evo-Eq}) with the generator
\begin{gather}
Z=\xi_1(x,t,u)\pde{\ }{x}+\xi_2(x,t,u)\pde{\ }{t}+\eta(x,t,u)\pde{\ }{u}
\end{gather}
is
\begin{gather}
\label{IC}
\left.
\vphantom{\frac{DA}{DB}}
L_E[u]Q\right|_{E=0}=0,
\end{gather}
where $Q$ denotes the corresponding invariant surface
\begin{gather}
\label{IF-Q}
Q=\eta(x,t,u)-\xi_1(x,t,u)u_x-\xi_2(x,t,u)u_t
\end{gather}
and
$L_E[u]$ denotes the linear operator
\begin{subequations}
\begin{gather}
\left.
L[u]=\pde{E}{u_t}D_t+\pde{E}{u}
+\pde{E}{u_x}D_x
+\pde{E}{u_{xx}}D_x^2
+\cdots+
\pde{E}{u_{nx}}D_x^n
\right|_{E:=u_t-F}
\\[0.3cm]
\qquad
=D_t-\pde{F}{u}
-\pde{F}{u_x}D_x
-\pde{F}{u_{xx}}D_x^2
-\cdots-
\pde{F}{u_{nx}}D_x^n.
\end{gather}
\end{subequations}
Here $D_t$ and $D_x$ denote the total $x$-derivative and total $t$-derivative, respectively.

Applying the invariance condition (\ref{IC}) for all equations in the sequence (\ref{gen-hier}) results in the following

\smallskip

\noindent
\begin{proposition}
\label{Prop-1}
For the general invariant surface (\ref{IF-Q}) of the sequence (\ref{gen-hier}), viz.
\begin{gather*}
u_t=\left(u_{(2k+1)x}\right)^{-n_k},\quad k=1,2,3,...,
\end{gather*}
we distinguish between two cases:

\smallskip

\smallskip

\noindent
{\bf Case 1:} For the sequence
\begin{gather}
\label{hier-1}
u_t=\left(u_{(2k+1)x}\right)^{-\frac{k}{k+1}},\quad k=1,2,3,...,
\end{gather}
the most general invariant surface is
\begin{subequations}
\begin{gather}
Q:=\left[
k(a_1+2a_2x)
+\left(\frac{k+1}{2k+1}\right)b_1\right]u
-u_x\left(a_0+a_1x+a_2x^2\right)\nn\\[0.3cm]
\qquad
-u_t\left(b_0+b_1t\right)+f(x),
\end{gather}
where
\begin{gather}
\frac{d^{2k+1}f(x)}{dx^{2k+1}}=0
\end{gather}
\end{subequations}
with $k=1,2,\ldots$. Here $a_0,\ a_1,\ a_2,\ b_0$ and $b_1$ are arbitrary constants. This gives the Lie symmetry algebra for 
(\ref{hier-1}) of dimension $2k+6$ which is spanned by the following set of Lie point symmetry generators:
\begin{gather}
\{Z_1=\pde{\ }{x},\quad Z_2=\pde{\ }{t},\quad Z_3=t\pde{\ }{t} +\left(\frac{k+1}{2k+1}\right)u\pde{\ }{u},\quad
Z_4=x\pde{\ }{x} +ku\pde{\ }{u},\nn\\[0.3cm]
\label{LA-1}
\quad
Z_5=x^2\pde{\ }{x} +2kxu\pde{\ }{u},\quad
Z_{p+5}=x^{p-1}\pde{\ }{u}\},\quad p=1,2,\ldots, 2k+1.
\end{gather} 

\smallskip 

\noindent
{\bf Case 2:} For the sequence
\begin{gather}
\label{gen-order-hier}
u_t=\left(u_{(2k+1)x}\right)^{-n_k}, \quad n_k\neq \frac{k}{k+1}, \quad k=1,2,3,...,
\end{gather} 
the most general invariant surface is
\begin{subequations}
\begin{gather}
Q:=\left(
\frac{a_1 n_k(2k+1)-b_1k}{n_k(k+1)-k}\right)u
-u_x\left(
\frac{(n_k+1)a_1-b_1}{n_k(k+1)-k}
+a_0
\right)\nn\\[0.3cm]
\qquad
-u_t\left(b_0+b_1t\right)+f(x),
\end{gather}
where
\begin{gather}
\frac{d^{2k+1}f(x)}{dx^{2k+1}}=0
\end{gather}
\end{subequations}
with $k=1,2,\ldots$. Here $a_0,\ a_1,\ b_0$ and $b_1$ are arbitrary constants. This gives the Lie symmetry algebra for 
(\ref{gen-order-hier}) of dimension $2k+5$ which is spanned by the following set of Lie point symmetry generators:
\begin{gather}
\{Z_1=\pde{\ }{x},\quad Z_2=\pde{\ }{t},\quad 
Z_3=x\pde{\ }{x}-\left[(k+1)n_k-k\right]t\pde{\ }{t}+ku\pde{\ }{u},\nn\\[0.3cm]
\label{LA-2}
\quad
Z_4=x\pde{\ }{x} +
\frac{(2k+1)n_k}{n_k+1}u\pde{\ }{u},\quad
Z_{p+4}=x^{p-1}\pde{\ }{u}\},\\[0.3cm]
\quad
p=1,2,\ldots, 2k+1.\nn
\end{gather}

\end{proposition}


\section{On the symmetry-integrability of the equations}

An evolution equation of order $n$ is said to be symmetry-integrable if it admits an infinite number of local Lie-Bäcklund symmetry generators
\begin{gather}
\label{LB}
Z_{LB}=Q(x,t,u,u_x,u_{xx},\ldots, u_{qx})\pde{\ }{u},
\end{gather}
where $q>n$ (for details, see for example \cite{Euler-book-2018}). The invariance condition to obtain Lie-Bäcklund symmetry generators is essentially the same as the condition (\ref{IC}), albeit
$Q$ now depends on derivatives up to some order $q$.
The first step is to establish the possible nonlinearities of the equations in the sequences of Proposition 1. 
For that we
conjecture a necessary condition which is based on many tedious calculations, the details of which we do not present here. To obtain Conjecture 1 (see below) we have considered the evolution equations in their most general form (\ref{Evo-Eq}) with $n\geq 2$
and applied the Lie-Bäcklund symmetry invariance condition to establish a necessary condition for the existence of 
Lie-Bäcklund symmetry generators of the form (\ref{LB}).
For this we have sought Lie-Bäcklund symmetries up to order 19. This leads to

\begin{conjecture}
\label{Conject-1}
The necessary condition for the symmetry-integrability of an evolution equation of the form
\begin{gather}
u_t=F(x,t,u,u_x,u_{xx},\ldots,u_{nx}),\quad n\geq 2,
\end{gather}
strictly depends on the order $n$ as follows:

\begin{enumerate}

\item
For $n=2$,  the necessary condition for the symmetry-integrability of the 2nd-order equation
\begin{subequations}
\begin{gather}
u_t=F(x,t,u,u_x,u_{xx})
\end{gather}
 is
\begin{gather}
\label{Conject-1}
2 \frac{\p F}{\p u_{xx}} \frac{\p^3 F}{\p u_{xx}^3}
-3\left(\frac{\p^2 F}{\p u_{xx}^2}\right)^2=0.
\end{gather}
\end{subequations}

\item
For $n=3$ ,the necessary condition for the symmetry-integrability of the 3rd-order equation
\begin{subequations}
\begin{gather}
u_t=F(x,t,u,u_x,u_{xx},u_{3x})
\end{gather}
 is
 \begin{gather}
\label{Cond-4th-order}
9\left(\frac{\p F}{\p u_{3x}}\right)^2
\frac{\p^4 F}{\p u_{3x}^4}
-45\frac{\p F}{\p u_{3x}}
\frac{\p^2 F}{\p u_{3x}^2}
\frac{\p^3 F}{\p u_{3x}^3}
+40\left(\frac{\p^2 F}{\p u_{3x}^2}\right)^3=0.
\end{gather}
\end{subequations}
\item
For odd $n\geq 5$, that is $n=(2k+3)$ with $k=1,2,3,...$, the necessary condition for the symmetry-integrability of 
\begin{subequations}
\begin{gather}
\label{case-2k+3}
u_t=F(x,t,u,u_x,u_{xx},\ldots, u_{2k+3}),\quad k=1,2,3,\ldots
\end{gather}
is
\begin{gather}
\label{cond-k-3rd-order}
(2k+3)\,\frac{\p F\ \ \  \ \ }{\p u_{(2k+3)x}}
\,
\frac{\p^3 F\ \ \  \ \ }{\p u_{(2k+3)x}^3}
-(3k+5)\left(
\frac{\p^2 F\ \ \  \ \ }{\p u_{(2k+3)x}^2}\right)^2
=0\\[0.3cm]
k=1,2,3,\ldots\ .\nn
\end{gather}
\end{subequations}

\item
For even $n\geq 4$, that is $n=2k+2$ with $k=1,2,3,\ldots$, the necessary condition for the symmetry-integrability of 
\begin{subequations}
\begin{gather}
\label{case-2k+3}
u_t=F(x,t,u,u_x,u_{xx},\ldots, u_{2k+2}),\quad k=1,2,3,\ldots
\end{gather}
is
\begin{gather}
\label{Conject-4}
\frac{\p^2 F\ \ \  \ \ }{\p u_{(2k+2)x}^2}=0,\quad k=1,2,3,\ldots\ .
\end{gather}
\end{subequations}

\end{enumerate}

\end{conjecture}

\strut\hfill

\noindent
{\bf Remarks:} {\it 
The necessary condition (\ref{Conject-1}) for 2nd-order evolution equations has been established in \cite{E-E-Conf-2024}. Note that (\ref{Conject-1}) is the Schwarzian derivative when we factor out 
$2\displaystyle{\left(\pde{F}{u_{xx}}\right)^2}$. Condition (\ref{Cond-4th-order}) for 3rd-order equations was derived 
in \cite{E-E-2019}.}

\strut\hfill

We now discuss the symmetry-integrability of the two sequences of Proposition 1 in view of Conjecture 1.

\strut\hfill

\noindent
{\bf Regarding Case 1:} We consider the sequence of equations (\ref{hier-1}), i.e.
\begin{subequations}
\begin{gather}
\label{Seq-1-a}
u_t=u_{3x}^{-1/2}\\[0.3cm]
\label{Seq-1-b}
u_t=u_{5x}^{-2/3}\\[0.3cm]
\label{Seq-1-c}
u_t=u_{7x}^{-3/4}\\[0.3cm]
\label{Seq-1-d}
u_t=u_{9x}^{-4/5}\\[0.3cm]
\mbox{etc.}\nn
\end{gather}
\end{subequations}
As reported in \cite{Euler-Euler-2025-Bluman}, equations (\ref{Seq-1-a}) and (\ref{Seq-1-b}) 
admit Lie-Bäcklund symmetry generators with lowest order 5 and 11, respectively, so that these two equations are symmetry-integrable. Moreover, equation (\ref{Seq-1-a}) satisfies condition (\ref{Cond-4th-order}) and equation (\ref{Seq-1-b}) satisfies condition (\ref{cond-k-3rd-order}) for $k=1$. 

It is easy to show that every equation of order $p\geq 5$ in the sequence (\ref{hier-1}) satisfies the necessary condition 
(\ref{cond-k-3rd-order})
for symmetry-integrability as given in Conjecture 1. In particular,
\begin{gather}
u_t=\left(u_{2k+1)x}\right)^{-\frac{k}{k+1}},\quad k=2,3,4,\ldots
\end{gather}
is equivalent to
\begin{gather}
u_t=\left(u_{(2k+3)x}\right)^{-\frac{k+1}{k+2}},\quad k=1,2,3,\ldots.
\end{gather}
whereby
\begin{gather}
F=\left(u_{(2k+3)x}\right)^{-\frac{k+1}{k+2}}
\end{gather}
satisfies condition (\ref{cond-k-3rd-order}) for $k=1,2,3,\ldots\ .$

Regarding the 7th-order equation (\ref{Seq-1-c}),
we found that this equation does not admit a Lie-Bäcklund symmetry generator up to order 19. Due to the memory restrictions of our computer we are not able to consider Lie-Bäcklund symmetry generators of order higher than 19, so we can therefore not make any statement about the existence of Lie-Bäcklund symmetries for the equations in sequence (\ref{hier-1}) that are of order 7 or higher.

\strut\hfill

\noindent
{\bf Regarding Case 2:} We consider the sequence of equations (\ref{gen-order-hier}), i.e.
\begin{subequations}
\begin{gather}
\label{Seq-2-a}
u_t=u_{3x}^{-n_1},\quad n_1\neq \frac{1}{2}\\[0.3cm]
\label{Seq-2-b}
u_t=u_{5x}^{-n_2},\quad n_2\neq \frac{2}{3}\\[0.3cm]
\label{Seq-2-c}
u_t=u_{7x}^{-n_3},\quad n_3\neq \frac{3}{4}\\[0.3cm]
\label{Seq-2-d}
u_t=u_{9x}^{-n_4},\quad n_4\neq \frac{4}{5}\\[0.3cm]
\mbox{etc.}\nn
\end{gather}
\end{subequations}
The 3rd-order equation (\ref{Seq-2-a}) satisfies condition (\ref{Cond-4th-order}) if and only if 
\begin{gather}
n_1\in\{-1,\,\frac{1}{2},\,2\}.
\end{gather} 
The only case of relevance here is $n_1=2$, i.e. 
\begin{gather}
\label{Seq-2-3rd-order}
u_t=u_{3x}^{-2},
\end{gather}
which has already been established as symmetry-integrable in \cite{E-E-81} (see equation (3.7) with $\alpha=0$ and $\beta=1$ in \cite{E-E-81}).

We now consider the equations of order $p\geq 5$ in the sequence (\ref{gen-order-hier}), namely
\begin{gather}
\label{Sec-13}
u_t=\left(u_{(2k+1)x}\right)^{-n_k},\quad n_k\neq \frac{k}{k+1},\quad k=2,3,4,\ldots\ ,
\end{gather}
which is equivalent to
\begin{gather}
u_t=\left(u_{(2k+3)x}\right)^{-n_{k+1}},\quad n_{k+1}\neq \frac{k+1}{k+2},\quad k=1,2,3,\ldots\ .
\end{gather}
Then
\begin{gather}
F=\left(u_{(2k+3)x}\right)^{-n_{k+1}}
\end{gather}
satisfies condition (\ref{cond-k-3rd-order}) if and only if
\begin{gather}
n_{k+1}\in\{-1,\ \frac{k+1}{k+2}\}
\end{gather}
for $k=1,2,3,\ldots\ .$ Therefore the sequence (\ref{Sec-13}) does not satisfy the necessary condition for 
symmetry-integrability for all $k=2,3,4,\ldots\ .$
\smallskip

Conjecture 1 now leads to the following

\strut\hfill

\noindent
{\bf Corollary 1.} {\it 
The sequence (\ref{gen-order-hier}), viz.
\begin{gather*}
u_t=\left(u_{(2k+1)x}\right)^{-n_k},\quad n_k\neq \frac{k}{k+1},\quad k=1,2,3,...,
\end{gather*} 
contains only one symmetry-integrable equation, namely the 3rd-order equation (\ref{Seq-2-3rd-order}), viz.
\begin{gather*}
u_t=u_{3x}^{-2}.
\end{gather*}
All remaining equations in this sequence are not symmetry-integrable.}

\section{Concluding remarks}

In this short Letter we have introduced two sequences of fully-nonlinear evolution equations, namely
\begin{gather}
\label{hier-1-Con}
u_t=\left(u_{(2k+1)x}\right)^{-\frac{k}{k+1}},\quad k=1,2,3,...
\end{gather}
and 
\begin{gather}
\label{hier-2-Con}
u_t=\left(u_{(2k+1)x}\right)^{-n_k},\quad n_k\neq \frac{k}{k+1},\quad k=1,2,3,...\ .
\end{gather} 
We report the Lie point symmetries of both sequences, where (\ref{hier-1-Con}) admits a Lie point symmetry algebra of dimension $2k+6$
spanned by the generators (\ref{LA-1}) and the sequence (\ref{hier-2-Con}) admits a Lie point symmetry algebra of dimension $2k+5$ spanned by the generators (\ref{LA-2}). The difference in dimensions of the two symmetry algebras is due to the Lie point symmetry generator 
\begin{gather}
\label{Z_5}
Z_5=x^2\pde{\ }{x}+2kxu\pde{\ }{u},
\end{gather}
which is admitted by the sequence (\ref{hier-1-Con}) but not by the sequence (\ref{hier-2-Con}).
Note that (\ref{Z_5}) generates the one-parameter local Lie transformation group
\begin{subequations}
\begin{gather}
\tilde x=\frac{x}{1-\epsilon x},\quad \tilde t=t,\quad
\tilde u=\frac{u}{(1-\epsilon x)^{2k}},
\end{gather}
\end{subequations}
where $\epsilon$ is the group parameter and $\tilde x(\epsilon=0)=x,\ \tilde t(\epsilon=0)=t,\ \tilde u(\epsilon=0)=u$.

Regarding the symmetry-integrability of the two sequences: We have established that every equation in the sequence
(\ref{hier-1-Con}) satisfies the necessary condition for symmetry-integrability. That is, condition (\ref{Cond-4th-order}) for the 3rd-order equation and condition (\ref{cond-k-3rd-order}) for the remaining equations in sequence 
(\ref{hier-1-Con}) which is based on Conjecture 1.
Furthermore, sequence (\ref{hier-1-Con}) contains at least two symmetry integrable equations, namely  \cite{Euler-Euler-2025-Bluman}
\begin{gather}
\label{3rd+5th}
u_t=u_{3x}^{-1/2}\quad \mbox{and} \quad u_t=u_{5x}^{-2/3}.
\end{gather}
As mentioned in Section 3, we are not able to find further symmetry-integrable equations of sequence
(\ref{hier-1-Con}) besides the two equations (\ref{3rd+5th}). It is therefore an open problem to find further symmetry-integrable equations or to prove that these two equations are the only symmetry-integrable equations in this sequence.

Regarding the sequence (\ref{hier-2-Con}): Based on Conjecture 1 we conclude in Corollary 1 that this series contains only one nonlinear symmetry-integrable equation, namely the 3rd-order equation (\ref{Seq-2-3rd-order}), {\it viz}
\begin{gather*}
u_t=u_{3x}^{-2}.
\end{gather*} 

Furthermore, by solving condition (\ref{cond-k-3rd-order}), Conjecture 1 leads to

\strut\hfill

\noindent
{\bf Corollary 2.} {\it Any fully-nonlinear evolution equations of dimension $1+1$ of order $n\geq 5$ that is symmetry-integrable has to be of the form
\begin{gather}
\label{conjecture-2}
u_t=f_1 \left(
\vphantom{\frac{DA}{DB}}
 u_{(2k+3)x}+f_2\right)^{-\frac{k+1}{k+2}}+f_3,\quad k=1,2,3,\ldots
\end{gather}
for some functions $f_j=f_j(x,t,u,u_x,u_{xx},\ldots,u_{(2k+2)x}),\ j=1,2,3$.
}

\strut\hfill

We hope that this preliminary results will pave the way for a deeper study of the sequences introduced here and possibly other sequences of (fully-)nonlinear evolution equations, in particular the sequence (\ref{conjecture-2}).

\label{lastpage}


\begin{thebibliography} {99}
\bibitem{Bluman+Anco}
Bluman, G W and Anco, S C, {\it Symmetry and Integration Methods for Differential Equations}, Springer, New York, 2002.

\bibitem{Euler-book-2018}
Euler M and Euler N, Nonlocal invariance of the multipotentialisations of the Kupershmidt equation and its higher-order hierarchies In: {\it Nonlinear Systems and Their Remarkable Mathematical Structures}, N Euler (ed), CRC Press, Boca Raton, 317-351, 2018 or {arXiv:2506.07780} [nlin.SI].

\bibitem{E-E-2019}
Euler M and Euler N, On Möbius-invariant and symmetry-integrable evolution equations
and the Schwarzian derivative, Studies in Applied Mathematics, 143(2), 139–156, 2019.

\bibitem{E-E-Conf-2024}
Euler M and Euler N, On 2nd-order fully-nonlinear equations with links to 3rd-order fully-nonlinear equations,
{\it Open Communications in Nonlinear Mathematical Physics}, {\bf vol. 4} (2), Proceedings of the OCNMP-2024 Conference, {\bf ocnmp:13765}, 158--170, 2024.
https://doi.org/10.46298/ocnmp.13765

\bibitem{Euler-Euler-2025-Bluman}
Euler M and Euler N, From fully-nonlinear to semilinear evolution equations: two symmetry-integrable examples,
{\it Open Communications in Nonlinear Mathematical Physics}, {\bf vol. 5}, Special Issue in honour of George W Bluman, {\bf ocnmp:15938}, 1--15, 2025.
https://doi.org/10.46298/ocnmp.15938



\bibitem{E-E-81}
Euler M and Euler N, On fully-nonlinear symmetry-integrable equations
with rational functions in their highest derivative: Recursion operators,
{\it Open Communications in Nonlinear Mathematical Physics}, {\bf 2}, {\bf ocnmp:10306}, 216--228, 2022. 
https://doi.org/10.46298/ocnmp.10306


\bibitem{Olver}
Olver PJ, {\it Applications of Lie Groups to Differential Equations}, Springer, New York, 1986.




\end {thebibliography}
\end{document}